\documentclass[reprint,longbibliography,amsmath,amssymb,aps,nofootinbib]{revtex4-1}
\usepackage{multirow} 
\usepackage{graphicx}
\usepackage{dcolumn}
\usepackage{bm}
\usepackage{amssymb,amsmath}
\usepackage{tabularx}
\usepackage{subcaption}
\usepackage{flushend}
\usepackage{lipsum}
\usepackage{float}
\usepackage{amsmath,multirow,bigdelim}
\usepackage{contour}
\usepackage{array}
\usepackage[bottom]{footmisc}
\usepackage{soul}
\usepackage{hyperref}
\newcolumntype{P}[1]{>{\centering\arraybackslash}p{#1}}
\usepackage{rotating} 
\usepackage{slashed}
\usepackage[normalem]{ulem}

\usepackage{amssymb}
\usepackage{pifont}
\newcommand{\beq}{\begin{equation}}
\newcommand{\eeq}{\end{equation}}
\newcommand{\bea}{\begin{eqnarray}}
\newcommand{\eea}{\end{eqnarray}}
\newcommand{\ba}{\begin{array}} 
\newcommand{\ea}{\end{array}}

\newcommand\gev{\,\mathrm{GeV}}
\newcommand\tev{\,\mathrm{TeV}}

\newcommand{\gsim}{\lower.7ex\hbox{$\;\stackrel{\textstyle>}{\sim}\;$}}
\newcommand{\lsim}{\lower.7ex\hbox{$\;\stackrel{\textstyle<}{\sim}\;$}}
\newcommand{\invab}{\,\mbox{ab}^{-1}}
\newcommand{\mll}{m_{\ell\ell}}

\definecolor{red}{rgb}{.9,0,0}
\definecolor{green}{rgb}{0,0.7,0}
\definecolor{blue}{rgb}{0,0,0.9}
\definecolor{orange}{rgb}{0.8,0.3,0}

\everymath{\displaystyle}

\begin{document}

\title{Limits on R-parity-violating couplings from Drell-Yan processes at the LHC}

\author{Saurabh Bansal$^a$}
\author{Antonio Delgado$^a$}
\author{Christopher Kolda$^a$}
\author{Mariano Quiros$^{a,b}$}
\affiliation{$^a$ Department of Physics, University of Notre Dame, 225 Nieuwland Hall, Notre Dame, Indiana 46556, USA\\ }

\affiliation{$^b$  Institut de F\'{\i}sica d'Altes Energies (IFAE) and BIST, Campus UAB\\ 08193, Bellaterra, Barcelona, Spain
}

\begin{abstract}
We find constraints on R-Parity Violating (RPV) couplings of the minimal supersymmetric standard model, using Drell-Yan differential cross sections at the LHC. Specifically, we look at the constraints on $\lambda'LQD^c$ couplings from monolepton and dilepton data published by ATLAS, with either electrons or muons in the final state. Out of the 18 RPV couplings to which the LHC is at least potentially sensitive by this technique, we find new limits on 12 (or 13) of them, for squarks masses above 1 (or 2)~TeV. We also show that one can employ our techniques to achieve significantly stronger bounds at a high-luminosity upgrade of the LHC.
\end{abstract}

\maketitle

\section{\label{section1}Introduction}

Supersymmetry (SUSY) has long been considered one of the best motivated extensions of the Standard Model (SM) and is therefore actively sought at the Large Hadron Collider (LHC) and in other experiments at both low and high energies. The absence of direct evidence for SUSY, however, has led physicists to consider regions of its parameter space where SUSY may remain hidden or just out of reach, including a reconsideration of the oft-made assumption of $R$-parity conservation. 

The advantages of $R$-parity are well known, including the elimination of a set of allowed terms in the Lagrangian that generate fast proton decay, and the prediction of a dark matter candidate whose production at the LHC plays a key role in most SUSY search strategies.
The imposition of $R$-parity, however, is not a requirement of SUSY.  It has long been understood that one could keep a subset of the terms forbidden by $R$-parity, either those that violate lepton number ($L$) or those that violate baryon number ($B$) -- but not both -- without destabilizing the proton.
Such models, known as $R$-parity-violating (RPV) models, have the disadvantage of not providing a candidate for the universe's dark matter, but are otherwise perfectly reasonable SUSY extensions of the SM. 
We will be studying here the version of RPV in which we allow those operators that violate $L$, specifically the superpotential operator $\lambda'_{ijk}L_iQ_jD^c_k$ since this operator can mediate tree-level quark to lepton processes by the exchange of squarks.

ATLAS and CMS have completed extensive searches for direct production of SUSY particles, both in $R$-parity conserving and violating modes, and have mass limits on sparticles typically in the 0.5 to 2.0~TeV range. However, the strongest quoted bounds on the RPV couplings are generally due to lower energy, high-precision measurements (see Ref.~\cite{Dercks:2017lfq} for a review). In this paper we will show that the LHC experiments now have enough data to place constraints on roughly half of the $\lambda'_{ijk}$ couplings that are stronger than those already in the literature. Furthermore, these bounds will continue to strengthen, by up to 4 times, as we move to a high-luminosity LHC.

Two ingredients are key to the results we show here. First is the observation that {\it tree-level}\/ sparticle exchange in RPV SUSY can generate processes that are identical in initial and final state to SM processes, and can therefore interfere with those SM process at the amplitude level; in $R$-parity-conserving SUSY, such an interference between SM and SUSY processes is only possible at loop level. The second ingredient is the observation that the presence of the operator $LQD^c$ in the superpotential causes the left-handed squarks (both up- and down-type) and the right-handed down-type squarks to behave as scalar leptoquarks. This allows them to couple to both initial state quarks and to final state leptons, mimicking Drell-Yan (DY) processes at the LHC. Moreover RPV SUSY models have been shown to contribute to processes like $b\to s\ell\ell$ and $b\to c\ell\nu_\ell$, and can potentially accommodate the lepton flavor universality violation which seems to appear in B-physics observables, such as $R_{D^{(\ast)}}$~\cite{Altmannshofer:2017poe} and/or $R_{K^{(\ast)}}$~\cite{Das:2017kfo}.

The effect of virtual exchange of scalar leptoquarks in DY processes has been studied previously~\cite{Raj:2016aky,Bansal:2018eha} and found to provide extremely strong constraints on the parameter space of those leptoquarks. In this paper, we extend those results to the study of RPV squarks and find similarly powerful constraints at the LHC. Furthermore, we show that with the accumulation of $3\invab$ of data, the LHC is sensitive to squark masses up to $20\tev$ for RPV couplings of $O(1)$. 

The rest of the paper is organized as follows: in section~\ref{section2} we will introduce the model and processes that will be studied; we will then present in section~\ref{section3} the analysis and in section~\ref{section4} the results; we will devote section~\ref{section5} to our conclusions.

\section{\label{section2}$L$-violating RPV in Drell-Yan Processes }

In the MSSM, the RPV portion of the superpotential can be written as,
$$
\mathcal{W} =\frac{1}{2} \lambda_{ijk} L_i L_j E^c_k+\lambda'_{ijk} L_i Q_j D^c_k+\frac{1}{2}\lambda''_{ijk} U^c_i D^c_j D^c_k.
$$
Here, $L$ and $Q$ are the $SU(2)$ doublets, while $E^c$, $D^c$ and $U^c$ are singlets. Using the standard notation, we have defined $\lambda_{ijk}$, $\lambda'_{ijk}$ and $\lambda''_{ijk}$ as new Yukawa couplings, where $i$, $j$ and $k$ are the generation indices; we omit a term that mixes sleptons and Higgs fields. If we enforce $B$-conservation, the $\lambda''$ are all zero, but the $\lambda$ and $\lambda'$ remain. The LHC is ill-suited to constrain the $\lambda$ couplings, as they generate dilepton-like interactions for the sleptons, interactions that are difficult to detect at a hadron collider. The $\lambda'$ interactions, on the other hand, cause the $\tilde Q$ and $\tilde D^c$ squarks to behave as leptoquarks, making them ideally suited for study at the LHC. 

The parameter space for the $\lambda'_{ijk}$ coupling is 27-dimensional, corresponding to all possible choices of $\{i, j, k\}$. Each of these couplings can modify the overall processes $pp\to\ell^+\ell^-$ and $pp\to\ell\bar\nu$, though the processes are much more sensitive to some couplings than to others, as we will discuss. In the following, we will consider, for the sake of simplicity, one element of $\lambda'_{ijk}$ at a time to be non-zero, while taking all other RPV couplings to be zero. 
Starting from the superpotential, we can then write the relevant pieces of the Lagrangian as:
\bea
\mathcal{L} &= \lambda'_{ijk} \Big[ \Big((V\bar{d}^c)_j P_L \nu_i - \bar{u}^c_j P_L \ell_i \Big) \tilde{d}_k^c + \Big(\bar d_k P_L \nu_i (V\tilde{d})_j-\nonumber\\
& \bar d_k P_L \ell_i \tilde{u}_j\Big)
+\left(\bar d_k P_L (V d)_j \tilde{\nu_i}+\bar d_k P_L u_j \tilde{\ell_i}\right)\Big].\nonumber
\eea 
Here $V$ is the CKM matrix; note, however, that the effect of including off-diagonal elements from $V$ in our analysis is quite small, usually less than a few percent. Moreover we work in the (excellent) approximation of massless neutrinos and ignore neutrino mixing effects.

\begin{figure}[!tbh]
	\begin{subfigure}[b]{0.235\textwidth}
		\includegraphics{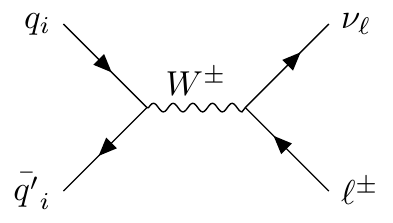} 
		\includegraphics{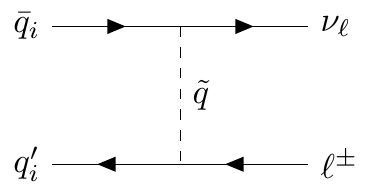} 
		\caption{}\label{FeynDiagA}
	\end{subfigure}
	\begin{subfigure}[b]{0.235\textwidth}
		\includegraphics{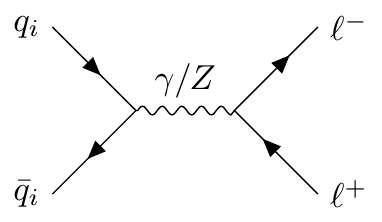} 
		\includegraphics{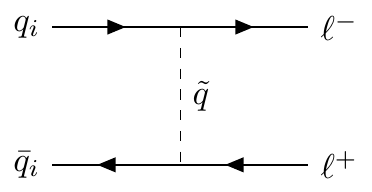} 
		\caption{}\label{FeynDiagB}
	\end{subfigure}
	\caption{\it \label{fig:FeynDiag}Feynman diagrams for the monolepton (\ref{FeynDiagA}) and dilepton (\ref{FeynDiagB}) production mediated by the SM gauge bosons (top) and RPV squarks (bottom).}
	\label{fig:Feyn}
\end{figure}

In this Lagrangian, the squarks are coupling to quarks and leptons in the manner of scalar leptoquarks, and so generate corrections to DY scattering as shown in Fig.~\ref{fig:Feyn}. More specifically, the Lagrangian above generates operators that contribute to DY scattering of the form (after Fierzing) of either $(\bar q_L\gamma^\mu q'_L)(\bar\ell_L \gamma_\mu \ell'_L)$ or $(\bar q_R\gamma^\mu q'_R)(\bar\ell_L \gamma_\mu \ell'_L)$, both of which interfere with pieces of the SM amplitudes. Because the leading effect comes through this interference, the contributions of the squarks to the differential cross sections come in proportional to $\lambda'^2/(t-m_{\tilde q}^2)$, rather than the fourth power, which partially accounts for the strength of our bounds.

Of course, not every choice of $\{i,j,k\}$ will provide a reasonable signal. For this study, we will only consider final state electrons and muons due to their clean reconstruction; thus we place no bounds on $\lambda'_{3jk}$ which leads to final state $\tau$ leptons. Furthermore, the initial state at the LHC is mostly $u$, $d$, and $s$-quarks for these purposes, and so the couplings $\lambda'_{ijk}$ with either $j$ or $k$ being 1 or 2 provide the strongest constraints. However, the couplings $\lambda'_{i3k}$ and $\lambda'_{ij3}$ can also be constrained for most choices of $j,k\neq 3$, since the 3rd generation squark can act as the mediator in these cases. Even the coupling $\lambda'_{i33}$ can be constrained by the dilepton data, though the constraint is not particularly strong due to its reliance on the tiny $b$-quark PDF. In all, we will find that our analysis of LHC data provides the strongest bounds for either 12 or 13 of the 18 couplings represented by $\lambda'_{ijk}$ and which involve either an electron or muon.

\section{\label{section3}Analysis }

Our analysis proceeds as follows. For each choice of $\{i,j,k\}$, we examine all relevant RPV contributions to $pp\to\ell^+\ell^-$ and $pp\to\ell\nu$, for $\ell=e,\mu$. In every case, the RPV contributions arise via the exchange of one or more squarks in the $t$-channel, as shown in Fig.~\ref{fig:FeynDiag}. If more than one squark contributes for a given choice of $\{i,j,k\}$, we take all the exchanged squarks to be degenerate. In order to obtain bounds on the RPV parameter space from current data, we compare our RPV model calculations, for $m_{\tilde q}\geq 1\tev$, against measurements published by ATLAS at $\sqrt{s} = 13\tev$ with 36~fb$^{-1}$ of integrated luminosity~\cite{1706.04786,1707.02424}.  Specifically, for the monolepton signals, we examine spectra of the monolepton transverse mass, $M_T \equiv [2 p^\ell_T \slashed{E}_T (1-\cos\Delta\phi)]^{1/2}$, where 
$p^\ell_T$ is the transverse momentum of the charged lepton,
$\slashed{E}_T$ is the missing transverse energy and
$\Delta\phi$ is the azimuthal opening angle between the two vectors. For the dileptons, we use the dilepton invariant mass, $\mll = \sqrt{\hat{s}}$. The analysis done here follows that of Refs.~\cite{Raj:2016aky,Bansal:2018eha} and more details can be found there.

Because our results hinge on the interference with the corresponding SM processes, we must generate event distributions in $M_T$ (for monoleptons) and $\mll$ (for dileptons) for the signal and the irreducible background together.
To that end, we employ the following procedure. 
Using the  MSTW 2008 NNLO PDFs~\cite{Martin:2009iq}, we first analytically calculate the SM and new physics mass spectra for $pp\to\ell^+\ell^-$ and $pp\to\ell\nu$ at the leading order. The resulting spectra are then rescaled by a global factor to account for the higher order corrections and lepton reconstruction efficiency, so that our background spectra match the irreducible backgrounds taken from ATLAS (and extracted via HEPData).
The net signal and background events are obtained by adding the reducible background extracted from Refs.~\cite{1706.04786,1707.02424} to our generated events.
To quantify the effect of our signals and estimate limits on RPV parameters, we use a very conservative version of a $\chi^2$ test that sums over all the ATLAS-defined bins, comparing both our model (with a given choice of $\lambda'_{ijk}$ and $m_{\tilde q}$) and the SM against current observations:
\beq
\chi^2_{\rm theory} = \sum_i^{\rm bins} \frac{(N_{{\rm theory}_i} - N_{{\rm data}_i})^2}{N_{{\rm data}_i}+\delta^2_{\rm sys}}~,
\label{eq:chisq}
\eeq
where $N_{{\rm theory}_i}$ is the number of events in bin $i$ predicted either by the SM or by the new physics and SM together. We then calculate a $\Delta\chi^2 = \chi^2_{\rm model} - \chi^2_{\rm SM}$;
the 95\% C.L.\ bound is located where $\Delta\chi^2$ is 5.99.
Here the systematic error $\delta_{\rm sys}$, taken to be a flat 6\% as seen across multiple bins in the ATLAS searches~\cite{1706.04786,1707.02424}, is assumed uncorrelated across bins, as the co-variance matrix has not been provided. Finally, we obtain our limits on the model by fitting a straight line in the plane $(m_{\tilde q},\lambda_{ijk}^\prime)$ to the 95\% C.L. contour in the region where $m_{\tilde q}\geq 1\tev$. The choice of region for the fit is based on the observation that, for squark masses $\geq 1\tev$, the 95\% C.L. contours become linear. And thus, the limits we quote only hold for the parameter space where $m_{\tilde{q}}>1\tev$. For lower squark masses, the bounds on $\lambda'_{ijk}$ depend on the squark mass in a more complicated way. However, for such light squarks one must also contend with strong direct production constraints.

Our analysis can be compared to one done entirely using 4-fermion effective operators, and in fact such an analysis could be reinterpreted to provide bounds on RPV couplings. However we find that the largest contributions to $\chi^2$ occur at $Q^2$ between roughly $500\gev$ and $1.5\tev$. Thus an effective operator approach is not appropriate except for squark masses well above $1\tev$. In comparing the results one obtains using effective operators instead of the full theory, we find that the effective operator approach typically generates bounds much stronger than one obtains using the full theory. For example, with the full propagator, we find that the current ATLAS monolepton data provide a bound: 
$$\lambda'_{111} <  0.16\frac{m_{\tilde d_R}}{\mbox{1 TeV}} + 0.049 ~\rm  (full) .$$ 
The same calculation with the effective operator approach results in a stronger limit of
$$ \lambda'_{111} <  0.16\frac{m_{\tilde d_R}}{\mbox{1 TeV}}~\rm (effective) .$$ 
Thus, for a down squark of mass 1~TeV, the effective operator method overestimates the bound on the coupling by about 30\%. But, as expected, the difference becomes negligible for much higher squark masses.

Finally, the limits we obtain are sensitive to current experimental uncertainties in the DY spectrum, and will therefore improve as the LHC accumulates more luminosity. In order to exploit this fact, we also make a simple projection of the expected limits on $\{m_{\tilde q}, \lambda'_{ijk}\}$ at the High Luminosity LHC with $3\invab$ of integrated luminosity and $\sqrt{s} = 13$ TeV. 
We calculate this reach by combining the monolepton and dilepton channels together, with a flat systematic error of 6\% and neglecting all sources of reducible background\footnote{This is different to what was done in Ref.~\cite{Bansal:2018eha}, where a flat 15\% error was used in the extrapolation to the High Luminosity LHC.}.

\section{\label{section4} Results}

Our results are summarized in Table~\ref{tab:i1} for $\lambda'_{1jk}$, and Table~\ref{tab:i2} for $\lambda'_{2jk}$; because we do not consider $\tau$ final states, we derive no bounds on $\lambda'_{3jk}$.  In the tables, we show first the strongest existing bound in the literature, collected and updated in Ref.~\cite{Barger:1989rk,ledroit:in2p3,Allanach:1999ic,Barbier:2004ez,Chemtob:2004xr,Dercks:2017lfq}. For $i=1$, those bounds are derived from: charged current universality for $\lambda'_{11k}$ and $\lambda'_{12k}$; atomic parity violation for $\lambda'_{131}$; the forward-backward asymmetry in $e^+e^-$ collisions for $\lambda'_{132}$; and bounds on neutrino masses for $\lambda'_{133}$ \cite{Godbole:1992fb}. For $i=2$, the bounds are derived from: $e-\mu$ universality in $\pi$-decays for $\lambda'_{21k}$; $\nu_\mu$ deep inelastic scattering for $\lambda'_{221}$ and $\lambda'_{231}$; from $D$-meson decays for $\lambda'_{222}$ and $\lambda'_{223}$; from $R_\mu = \Gamma(Z\to \textrm{had})/\Gamma(Z\to \mu \bar \mu)$ for $\lambda'_{232}$ \cite{Bhattacharyya:1995pr}; and again from bounds on neutrino masses for $\lambda'_{233}$ \cite{Godbole:1992fb}.

The third (fourth) column in the tables indicates the limits obtained using the monolepton (dilepton) data from ATLAS, and the last column indicates the expected limits from a $3\invab$ high luminosity LHC. The strongest constraint on a particular coupling for $m_{\tilde{q}} \geq 1\tev$ has been highlighted by showing it in a box. For the case of $\lambda'_{121}$, we show two boxes because the lines for these two constraints intersect, with the dilepton constraint becoming the dominant one for $m_{\tilde{q}} \gsim 2\tev$.

An interesting feature in Tables~\ref{tab:i1}-\ref{tab:i2} is that the monolepton constraints on $\lambda'_{2jk}$ are much stronger than those on $\lambda'_{1jk}$, and the current monolepton constraints on $\lambda'_{2jk}$ are not all that much weaker than we project for the $3\invab$ LHC. Both of these results can be understood by observing that the current ATLAS monolepton data for muons show a small {\it excess}\/ over the SM predictions for most of the $m_T$-bins, while the RPV monolepton operator interferes {\it destructively}\/ with the SM, pulling down the expected cross section. Thus the resulting constraints on $\lambda'_{2jk}$ are much stronger than one would expect just by comparing to the SM distribution. This same excess of muon events in the data also results in a constraint on $\lambda'_{22k}$ that is surprisingly strong despite being suppressed by the $c$-quark PDF. Thus one should keep in mind that if the mono-muon data were to become more closely aligned with the SM prediction, the bounds on $\lambda'_{2jk}$ from mono-muons would weaken, typically to match those derived on $\lambda'_{1jk}$ from mono-electrons.
This is the case we consider in our projection for the $3\invab$ limits, in that we assume that the data will match the SM expectations.

\begin{table*}[t]
	\begin{ruledtabular}
		{\renewcommand{\arraystretch}{2.5}
			\begin{tabular}{c c c c c}
				$ijk$ 
				&
				Literature & Monolepton & Dilepton & 
				Projected \\
				\hline \hline
				111 & $0.21  \frac{m_{\tilde{d}_R}}{1 \tev}$ & \fbox{$0.16 \frac{m_{\tilde{d}_R}}{1 \tev}+0.030$} & $0.31 \frac{m_{\tilde{q}}}{1 \tev}+0.14$ & $0.049 \frac{m_{\tilde{q}}}{1 \tev}+0.023$ \\ 
				112 & $0.21  \frac{m_{\tilde{s}_R}}{1 \tev}$ & \fbox{$0.16 \frac{m_{\tilde{s}_R}}{1 \tev}+0.030$} & $0.30 \frac{m_{\tilde{q}}}{1 \tev}+0.15$ & $0.053 \frac{m_{\tilde{q}}}{1 \tev}+0.020$ \\ 
				113 & $0.21  \frac{m_{\tilde{b}_R}}{1 \tev}$ & \fbox{$0.16 \frac{m_{\tilde{b}_R}}{1 \tev}+0.030$} & $0.29 \frac{m_{\tilde{q}}}{1 \tev}+0.14$ & $0.053 \frac{m_{\tilde{q}}}{1 \tev}+0.020$ \\

				121 & \fbox{$0.43  \frac{m_{\tilde{d}_R}}{1 \tev}$} & $0.70 \frac{m_{\tilde{d}_R}}{1 \tev}+0.41$ & \fbox{$0.34 \frac{m_{\tilde{q}}}{1 \tev}+0.18$} & $0.076 \frac{m_{\tilde{q}}}{1 \tev}+0.028$ \\ 
				122 & \fbox{$0.43  \frac{m_{\tilde{s}_R}}{1 \tev}$} & $0.70 \frac{m_{\tilde{s}_R}}{1 \tev}+0.41$ & $0.48 \frac{m_{\tilde{q}}}{1 \tev}+0.28$ & $0.34 \frac{m_{\tilde{q}}}{1 \tev}+0.24$ \\ 
				123 & \fbox{$0.43  \frac{m_{\tilde{b}_R}}{1 \tev}$} & $0.70 \frac{m_{\tilde{b}_R}}{1 \tev}+0.41$ & $0.49 \frac{m_{\tilde{q}}}{1 \tev}+0.28$ & $0.36 \frac{m_{\tilde{q}}}{1 \tev}+0.22$ \\ 
				
				131 & \fbox{$0.19  \frac{m_{\tilde{t}_L}}{1 \tev}$} & $-$ & $0.34 \frac{m_{\tilde{q}}}{1 \tev}+0.16$ & $0.075 \frac{m_{\tilde{q}}}{1 \tev}+0.036$ \\ 
				132 & $2.8  \frac{m_{\tilde{t}_L}}{1 \tev}$ & $-$ & \fbox{$0.60 \frac{m_{\tilde{q}}}{1 \tev}+0.37$} & $0.43 \frac{m_{\tilde{q}}}{1 \tev}+0.48$  \\  
				133 & \fbox{$0.0044  \sqrt{\frac{m_{\tilde{b}}}{1 \tev}}$} & $-$ & $0.72 \frac{m_{\tilde{q}}}{1 \tev}+0.46$ & $0.57 \frac{m_{\tilde{q}}}{1 \tev}+0.55$ 
			
			\end{tabular}
		}
	\end{ruledtabular}
	\caption{\it \label{tab:i1}%
		Upper bounds on $\lambda'_{1jk}$ from the literature and derived in this study. The strongest current constraint on a particular coupling (for $m_{\tilde{q}} > 1 \tev$) is shown in a box. For the dilepton bounds, $\tilde q$ represents $\tilde u_{j,L}$ and $\tilde d_{k,R}$, taken degenerate. Projected bounds assume $3\invab$ of data, combining both mono- and dilepton analyses where appropriate.}
\end{table*}

\begin{table*}[t]
	\begin{ruledtabular}
		{\renewcommand{\arraystretch}{2.5}
			\begin{tabular}{c c c c c}
				$ijk$ of $\lambda'_{ijk}$&
				Literature & Monolepton & Dilepton & 
				Projected \\
				\hline \hline
				211 & $0.59  \frac{m_{\tilde{d}_R}}{1 \tev}$ & \fbox{$0.090 \frac{m_{\tilde{d}_R}}{1 \tev}+0.014$} & $0.31 \frac{m_{\tilde{q}}}{1 \tev}+0.098$ & $0.050 \frac{m_{\tilde{q}}}{1 \tev}+0.027$ \\ 
				212 & $0.59  \frac{m_{\tilde{s}_R}}{1 \tev}$ & \fbox{$0.090 \frac{m_{\tilde{s}_R}}{1 \tev}+0.014$} & $0.33 \frac{m_{\tilde{q}}}{1 \tev}+0.20$ & $0.053 \frac{m_{\tilde{q}}}{1 \tev}+0.028$ \\ 
				213 & $0.59  \frac{m_{\tilde{b}_R}}{1 \tev}$ & \fbox{$0.090 \frac{m_{\tilde{b}_R}}{1 \tev}+0.014$} & $0.33 \frac{m_{\tilde{q}}}{1 \tev}+0.19$ & $0.053 \frac{m_{\tilde{q}}}{1 \tev}+0.029$ \\

				221 & $1.8  \frac{m_{\tilde{s}_R}}{1 \tev}$ & $0.44 \frac{m_{\tilde{d}_R}}{1 \tev}+0.040$ & \fbox{$0.34 \frac{m_{\tilde{q}}}{1 \tev}+0.074$} & $0.080 \frac{m_{\tilde{q}}}{1 \tev}+0.036$ \\ 
				222 & $2.1  \frac{m_{\tilde{s}_R}}{1 \tev}$ & \fbox{$0.44 \frac{m_{\tilde{s}_R}}{1 \tev}+0.040$} & $0.57 \frac{m_{\tilde{q}}}{1 \tev}+0.33$ & $0.35 \frac{m_{\tilde{q}}}{1 \tev}+0.27$ \\ 
				223 & $2.1  \frac{m_{\tilde{b}_R}}{1 \tev}$ & \fbox{$0.44 \frac{m_{\tilde{b}_R}}{1 \tev}+0.040$} & $0.59 \frac{m_{\tilde{q}}}{1 \tev}+0.33$ & $0.37 \frac{m_{\tilde{q}}}{1 \tev}+0.25$ \\ 
				
				231 & $1.8  \frac{m_{\tilde{b}_L}}{1 \tev}$ & $-$ & \fbox{$0.34 \frac{m_{\tilde{q}}}{1 \tev}+0.074$} & $0.081 \frac{m_{\tilde{q}}}{1 \tev}+0.033$ \\ 
				232 & $0.56  $ & $-$ &\fbox{$0.70 \frac{m_{\tilde{q}}}{1 \tev}+0.42$} & $0.45 \frac{m_{\tilde{q}}}{1 \tev}+0.48$ \\ 
				233 & \fbox{$0.47  \sqrt{\frac{m_{\tilde{b}}}{1 \tev}}$} & $-$ & $0.90 \frac{m_{\tilde{q}}}{1 \tev}+0.49$ & $0.60 \frac{m_{\tilde{q}}}{1 \tev}+0.54$ 
			\end{tabular}
		}
	\end{ruledtabular}
	\caption{\it \label{tab:i2}%
		Upper bounds on $\lambda'_{2jk}$. See Table~\ref{tab:i1} caption for details. The bound on $\lambda'_{232}$ assumes $m_{\tilde q} = 100\gev$.}
\end{table*}

\section{\label{section5} Conclusions}
In this paper we found that one can use the monolepton and dilepton DY data from ATLAS to derive bounds on a total of 12 (or 13) RPV couplings that are stronger -- sometimes far stronger -- than any found in the literature, assuming squark masses above 1 (or 2)~TeV. In particular, roughly half of the $\lambda'$ couplings involving electrons, and all but one of the couplings involving muons, are better constrained by the LHC than by any other existing experimental constraint. These constraints can also be interpreted as bounds on squark masses, if one assumes some value for the RPV couplings. For example, if we assume $\lambda'_{111} = 1$, we find that $m_{\tilde d_R} > 6.1\tev$, far stronger than direct search bounds. 

Furthermore, because these constraints are all statistics limited, they will improve significantly with additional luminosity. For example, the bounds on $\lambda'$ from electrons will typically improve by a factor of 2.5--4 by accumulating $3\invab$ of data at $\sqrt{s} = 13\tev$; the corresponding bounds on squark masses for $\mathcal O(1)$ couplings are in the 20~TeV range. We therefore strongly encourage the LHC collaborations to reanalyze their DY data in the context of RPV supersymmetry in order to obtain the most trustworthy bounds possible. 

\section*{Acknowledgments}

We will like to thank useful conversations with Nirmal Raj. This works was partially supported by the National Science Foundation under grant PHY-1820860.  The work of MQ is partly supported by Spanish MINEICO (grants CICYT-FEDER-FPA2014-55613-P and FPA2017-88915-P), by the Catalan Government under grant 2017SGR1069, and Severo Ochoa Excellence Program of MINEICO (grant SEV-2016-0588).

\end{document}